\documentclass[twocolumn]{aastex631}

\usepackage{lineno}

\newcommand{\cenDM}{20.39}
\newcommand{\cenDMerr}{0.08}
\newcommand{\cenDKPC}{119.8}
\newcommand{\cenDKPCerr}{4.1}
\newcommand{\cenRA}{189.5908}
\newcommand{\cenRAerr}{9.5\arcsec}
\newcommand{\cenDEC}{-40.9043}
\newcommand{\cenDECerr}{10.5\arcsec}
\newcommand{\cenELLIP}{0.36}
\newcommand{\cenELLIPerr}{0.05}
\newcommand{\cenRH}{2.60}
\newcommand{\cenRHerr}{0.30}
\newcommand{\cenRHPHYS}{90.6}
\newcommand{\cenRHPHYSerr}{11}
\newcommand{\cenRGEO}{2.08}
\newcommand{\cenRGEOerr}{0.25}
\newcommand{\cenRGEOphys}{72.6}
\newcommand{\cenRGEOphyserr}{9.2}
\newcommand{\cenTHETA}{10}
\newcommand{\cenTHETAerr}{4}
\newcommand{\cenMV}{-5.39}
\newcommand{\cenMVerr}{0.19}
\newcommand{\cenPMRA}{-0.14}
\newcommand{\cenPMRAerr}{0.05}
\newcommand{\cenPMDEC}{-0.19}
\newcommand{\cenPMDECerr}{0.04}

\newcommand{\eriDM}{19.22}
\newcommand{\eriDMerr}{0.11}
\newcommand{\eriDKPC}{69.9}
\newcommand{\eriDKPCerr}{3.6}
\newcommand{\eriRA}{76.4246}
\newcommand{\eriRAerr}{16.4\arcsec}
\newcommand{\eriDEC}{-9.5189}
\newcommand{\eriDECerr}{15.3\arcsec}
\newcommand{\eriELLIP}{0.26}
\newcommand{\eriELLIPerr}{0.09}
\newcommand{\eriRH}{3.24}
\newcommand{\eriRHerr}{0.48}
\newcommand{\eriRHPHYS}{65.9}
\newcommand{\eriRHPHYSerr}{10}
\newcommand{\eriRGEO}{2.78}
\newcommand{\eriRGEOerr}{0.45}
\newcommand{\eriRGEOphys}{56.5}
\newcommand{\eriRGEOphyserr}{9.5}
\newcommand{\eriTHETA}{75}
\newcommand{\eriTHETAerr}{26}
\newcommand{\eriMV}{-3.55}
\newcommand{\eriMVerr}{0.24}
\newcommand{\eriPMRA}{0.22}
\newcommand{\eriPMRAerr}{0.06}
\newcommand{\eriPMDEC}{-0.11}
\newcommand{\eriPMDECerr}{0.05}

\usepackage{threeparttable}
\usepackage{xspace}
\begin{document}

\shortauthors{Casey et al.}
\shorttitle{Cen~I and Eri~IV}

\title{Deep Photometric Observations of Ultra-Faint Milky Way Satellites Centaurus I and Eridanus IV\footnote{This paper includes data gathered with the 6.5 m Magellan Telescopes at Las Campanas Observatory, Chile.}}

\author[0009-0005-9002-4800]{Quinn O. Casey}
\affiliation{Department of Physics and Astronomy, Dartmouth College, Hanover, NH 03755, USA}

\author[0000-0001-9649-4815]{Bur\c{c}in Mutlu-Pakdil}
\affil{Department of Physics and Astronomy, Dartmouth College, Hanover, NH 03755, USA}

\author[0000-0003-4102-380X]{David J. Sand}
\affiliation{Steward Observatory, University of Arizona, 933 North Cherry Avenue, Tucson, AZ 85721-0065, USA}

\author[0000-0002-6021-8760]{Andrew B. Pace}
\affil{Department of Astronomy, University of Virginia, 530 McCormick Road, Charlottesville, VA 22904, USA}

\author[0000-0002-1763-4128]{Denija Crnojevi\'{c}}
\affil{Department of Physics and Astronomy, University of Tampa, 401 West Kennedy Boulevard, Tampa, FL 33606, USA}

\author[0000-0001-9775-9029]{Amandine Doliva-Dolinsky}
\affil{Department of Physics and Astronomy, Dartmouth College, Hanover, NH 03755, USA}

\author[0000-0003-1697-7062]{William Cerny}
\affil{Department of Astronomy, Yale University, New Haven, CT 06520, USA}

\author[0000-0002-2446-8332]{Mairead E. Heiger}
\affil{Department of Astronomy and Astrophysics, University of Toronto, 50 St. George Street, Toronto ON, M5S 3H4, Canada}
\affil{Dunlap Institute for Astronomy and Astrophysics, University of Toronto, 50 St. George Street, Toronto ON, M5S 3H4, Canada}

\author[0000-0001-5805-5766]{Alex H. Riley}
\affil{Institute for Computational Cosmology, Department of Physics, Durham University, South Road, Durham DH1 3LE, UK}

\author[0000-0002-4863-8842]{Alexander P. Ji}
\affil{Department of Astronomy \& Astrophysics, University of Chicago, 5640 S Ellis Avenue, Chicago, IL 60637, USA}
\affil{Kavli Institute for Cosmological Physics, University of Chicago, Chicago, IL 60637, USA}

\author[0000-0002-9269-8287]{Guilherme Limberg}
\affil{Kavli Institute for Cosmological Physics, University of Chicago, Chicago, IL 60637, USA}
\affil{Universidade de S\~ao Paulo, IAG, Departamento de Astronomia, SP 05508-090, S\~ao Paulo, Brazil}

\author[0009-0008-3389-9848]{Laurella Marin}
\affiliation{Department of Physics and Astronomy, Dartmouth College, Hanover, NH 03755, USA}

\author[0000-0002-9144-7726]{Clara E. Martínez-Vázquez}
\affil{International Gemini Observatory/NSF NOIRLab, 670 N. A'ohoku Place, Hilo, Hawai'i, 96720, USA}

\author[0000-0003-0105-9576]{Gustavo E. Medina}
\affil{Department of Astronomy and Astrophysics, University of Toronto, 50 St. George Street, Toronto ON, M5S 3H4, Canada}

\author[0000-0002-9110-6163]{Ting S. Li}
\affil{Department of Astronomy and Astrophysics, University of Toronto, 50 St. George Street, Toronto ON, M5S 3H4, Canada}

\author[0009-0007-9488-7050]{Sasha N. Campana}
\affil{Department of Physics and Astronomy, Dartmouth College, Hanover, NH 03755, USA}

\author[0000-0001-5143-1255]{Astha Chaturvedi}
\affil{Department of Physics, University of Surrey, Guildford GU2 7XH, UK}

\author[0000-0002-1594-1466]{Joanna D. Sakowska}
\affil{Department of Physics, University of Surrey, Guildford GU2 7XH, UK}

\author[0000-0001-6455-9135]{Alfredo Zenteno}
\affil{Cerro Tololo Inter-American Observatory/NSF NOIRLab, Casilla 603, La Serena, Chile}

\author[0000-0002-3690-105X]{Julio A. Carballo-Bello}
\affil{Instituto de Alta Investigaci\'on, Universidad de Tarapac\'a, Casilla 7D, Arica, Chile}

\author[0000-0001-9438-5228]{Mahdieh Navabi}
\affil{Department of Physics, University of Surrey, Guildford GU2 7XH, UK}

\author[0000-0003-4383-2969]{Clecio R. Bom}
\affil{Centro Brasileiro de Pesquisas F\'isicas, Rua Dr. Xavier Sigaud 150, 22290-180 Rio de Janeiro, RJ, Brazil}

\collaboration{22}{(DELVE Collaboration)}

\correspondingauthor{Quinn Casey}
\email{quinn.o.casey.gr@dartmouth.edu}

\begin{abstract}
We present deep Magellan$+$Megacam imaging of Centaurus~I (Cen~I) and Eridanus~IV (Eri~IV), two recently discovered Milky Way ultra-faint satellites. 
Our data reach $\sim2-3$ magnitudes deeper than the discovery data from the DECam Local Volume Exploration (DELVE) Survey.
We use these data to constrain their distances, structural properties (e.g., half-light radii, ellipticity, and position angle), and luminosities. 
We investigate whether these systems show signs of tidal disturbance, and identify new potential member stars using \textit{Gaia} EDR3.
Our deep color-magnitude diagrams show that Cen~I and Eri~IV are consistent with an old ($\tau\sim 13.0$~Gyr) and metal-poor ($\text{[Fe/H]}\le-2.2$) stellar population. 
We find Cen~I to have a half-light radius of $r_{h}=\cenRH\pm\cenRHerr\arcmin$ ($\cenRHPHYS \pm\cenRHPHYSerr$ pc), an ellipticity of $\epsilon=\cenELLIP\pm\cenELLIPerr$, a distance of $D=\cenDKPC\pm\cenDKPCerr$ kpc ($m-M=\cenDM\pm\cenDMerr$ mag), and an absolute magnitude of $M_{V}=\cenMV\pm\cenMVerr$. 
Similarly, Eri~IV has $r_{h}=\eriRH\pm\eriRHerr\arcmin$ ($\eriRHPHYS\pm\eriRHPHYSerr$ pc), $\epsilon=\eriELLIP\pm\eriELLIPerr$, $D=\eriDKPC\pm\eriDKPCerr$ kpc ($m-M=\eriDM\pm\eriDMerr$ mag), and $M_{V}=\eriMV\pm\eriMVerr$. 
These systems occupy a space on the size-luminosity plane consistent with other known Milky Way dwarf galaxies which supports the findings from our previous spectroscopic follow-up.
Cen~I has a well-defined morphology which lacks any clear evidence of tidal disruption, whereas Eri~IV hosts a significant extended feature with multiple possible interpretations.

\end{abstract}


\section{Introduction} \label{sec:introduction}
Ultra-faint dwarfs (UFDs) are the oldest ($\tau \gtrsim 13.0$~Gyr), faintest ($L \lesssim 10^{5}L_{\odot}$, $M_{V}\gtrsim-7.7$), least massive ($M \lesssim 10^{5}M_{\odot}$), most metal-poor, and most dark matter dominated galactic systems known in the Universe \citep{Simon2019}. 
Therefore, they can be used as a unique laboratory to test the nature of dark matter, the validity of Lambda cold dark matter ($\Lambda$CDM), and galaxy formation on the smallest scales \citep[e.g.][]{Weinberg2015, Gonzalez2017, Bullock2017, Strigari2018, Safarzadeh2020}. 

Over the last decade, there has been significant progress in the discovery space of UFDs around the Milky Way \citep[MW; see][and references therein]{Simon2019}. 
However, most newly discovered systems reside at the very limit of the survey data in which they were discovered, and their true nature remain uncertain, partially due to the lack of deep and wide-field imaging. 
New ultra-faint satellites only have a handful of detectable stars in the discovery data with which to infer their properties. 
Therefore, it is imperative to follow up new systems to derive robust measurements of their structural parameters, distances, luminosities, as well as understanding whether a system is clearly disrupting. 
These findings should then be interpreted in the context of known UFD galaxies.

In this paper we focus on Centaurus~I (Cen~I) and Eridanus~IV (Eri~IV), which were discovered in the DECam Local Volume Exploration survey (DELVE\footnote{\url{https://delve-survey.github.io/}}; Cen~I by \citealt{Mau2020}, and Eri~IV by \citealt{Cerny2021}). 
Both systems were identified using the {\fontfamily{cmtt}\selectfont simple}\footnote{\url{https://github.com/DarkEnergySurvey/simple}} algorithm. 
This algorithm searches for local spatial over-densities consistent with old and metal-poor stellar populations, and has been used to discover over 30 MW satellites \citep[e.g.][]{Bechtol2015, DW2015, Mau2019, Mau2020, Cerny2021, Cerny2023a, Cerny2023b, Cerny2023c, McNanna2024, Tan2024, Cerny2024}. 
\citet{Martinez-Vazquez2021} detected three RR Lyrae stars in Cen~I and used them to determine a distance of $D=117.7\pm0.1$ kpc ($\pm4$ kpc systematic error).
\citet{Heiger2023} observed 34 member stars of Cen~I and 28 member stars of Eri~IV  using the Inamori Magellan Areal Camera \& Spectrograph (IMACS)/Magellan. 
They measured the velocity and metallicity dispersions ($\sigma_v=4.2^{+0.6}_{-0.5}$~km~s$^{-1}$, $\sigma_{\text{[Fe/H]}}=0.38^{+0.07}_{-0.05}$ for Cen~I; $\sigma_v=6.1^{+1.2}_{-0.9}$~km~s$^{-1}$, $\sigma_{\text{[Fe/H]}}=0.20\pm0.09$ for Eri~IV), and concluded that the systems are dark-matter-dominated and exhibit properties largely consistent with other known UFDs. 


Interestingly, both systems show tentative signs of tidal disruption, manifested by nearby stellar overdensities in the discovery data. It is essential to investigate these features with deeper follow-up observations, as including unbound stars in dynamical analyses could result in overestimated dark matter masses, and theoretical models attempting to replicate dwarf properties would be based on incorrect luminosities. Here, we examine their outer structures, search for extremely low surface brightness extensions, and assess whether these systems are in dynamical equilibrium.

The paper is organized as follows: in Section \ref{sec:data}, we describe the observations and the data reduction. Section \ref{sec:analysis} discusses our analysis of the data regarding the systems' distances, their structural properties, their absolute magnitudes, the presence of any extended structures, and we search for potential new member stars in \textit{Gaia}. We discuss the results of our analysis and their importance in Section \ref{sec:discussion}. Finally, we summarize our main results in Section \ref{sec:conclusion}. 

\section{Observations and Data Reduction} \label{sec:data}
Cen~I and Eri~IV were observed in the $g$ and $r$ bands using Megacam at the $f$/5 focus on the Magellan Clay telescope. 
The data for Cen~I and Eri~IV were taken on January 28, 2022, and February 2, 2022, respectively. 
Magellan/Megacam uses 36 2048$\times$4608 pixel CCDs, each with a scale of 0.08\arcsec/pixel (which were binned $2\times2$). 
This yields a field of view (FoV) of 24\arcmin$\times$24\arcmin \space \citep{McLeod2015}. 
We obtained $7\times300$s exposures in $g$ and $r$ for Cen~I, and $6\times300$s in $g$ and $r$ for Eri~IV (see Table~\ref{table:1}). 
The data were reduced using the Megacam pipeline developed at the Harvard-Smithsonian Center for Astrophysics by M. Conroy, J. Roll, and B. McLeod; this process includes detrending the data and performing astrometry. We then stack the individual dithered frames with {\fontfamily{cmtt}\selectfont SWarp} \citep{Bertin2002}.

We perform point-spread function (PSF) photometry using the DAOPHOTII/ALLSTAR software \citep{Stetson1994}, and follow the methodology detailed in \citet{Burcin2018}. 
We run ALLSTAR twice: first on the final stacked image, then again on the final stacked image after subtracting the stars found in the first run.
This methodology allows us to recover much fainter sources. 
We cull our catalog by removing objects which are not point sources; i.e., we remove outliers in $\chi^{2}$ vs. magnitude, magnitude error vs. magnitude, and sharpness vs. magnitude. 
We positionally match our source catalogs derived from the $g$- and $r$-band images using a maximum matching radius of 0.5\arcsec. 
Only those point sources detected in both bands are used to create our final catalog for both systems.

We calibrate the output of our stellar photometry by matching with DELVE data release 2 \citep[DR2, ][]{delvedr2}. We use all stars within the FoV where 17.5 $< g <$ 21 and 17.5 $< r <$ 21 for calibration. We correct for Galactic extinction on a star-by-star basis using the \citet{Schlegel1998} dust maps; the average $E(B-V)$ values are 0.127 for Cen~I and 0.109 for Eri~IV. All quoted magnitudes throughout this paper are the extinction corrected values. Tables \ref{tab:cen_table} and \ref{tab:eri_table} show the photometric catalogs containing the calibrated magnitudes (uncorrected for extinction), DAOPHOT uncertainties, and Galactic extinction for Cen~I and Eri~IV, respectively.

We use the DAOPHOT ADDSTAR routine to inject artificial stars into the data; this allows us to determine our photometric uncertainties and completeness as a function of magnitude and color. Following \citet{Burcin2018}, we inject artificial stars into our images on a grid. The $r$ band magnitudes range from 18--29~mag, where the values take on fainter magnitudes with an exponentially increasing probability. The $g$ band magnitudes are randomly selected based on the $g-r$ color ranging from $-0.5$ to $1.5$ mag. We perform the artificial star injections 10 different times, each iteration inserts $\sim$100,000 artificial stars in each field. We then run DAOPHOT and ALLSTAR twice on the images containing the artificial stars in the exact same manner used on the real data. We require the same point-source selection criteria on $\chi^{2}$, magnitude error, and sharpness as was required on the real data. We use these artificial star catalogs to determine our 50\% and 90\% completeness (see Table~\ref{table:1}) and photometric uncertainties. 

\begin{table}[t]
\centering
\begin{tabular*}{\linewidth}{c c c c c c c}
 \hline \hline
 Dwarf & UT Date & Filter & Exp & 50\% & 90\% \\ 
  Name & & & (s) & (mag) & (mag) \\
 \hline
  Cen~I & 2022 Jan 28 & $g$ & 7$\times$300 & 27.43 & 25.89 \\
   & 2022 Jan 28 & $r$ & 7$\times$300 & 26.92 & 25.16 \\
  Eri~IV & 2022 Feb 2 & $g$ & 6$\times$300 & 25.37 & 24.11 \\
   & 2022 Feb 2 & $r$ & 6$\times$300 & 25.02 & 23.45 \\
 \hline
\end{tabular*}
\caption{Summary of Magellan/Megacam observations and field completeness.}
\label{table:1}
\end{table}

\begin{table*}[t]
\caption{Cen~I Photometry in the DELVE Photometric System}
    \begin{tabular}{ccccccccc}
    \hline \hline
        Star No. & $\alpha$ & $\delta$ & $g$ & $\delta g$ & $A_{g}$ & $r$ & $\delta r$ & $A_{r}$ \\
         & (deg J2000.0) & (deg J2000.0) & (mag) & (mag) & (mag) & (mag) & (mag) & (mag) \\
        \hline
        0 & 189.40289 & -41.054473 & 19.360 & 0.003 & 0.334 & 18.449 & 0.003 & 0.225 \\
        1 & 189.37668 & -40.915770 & 19.528 & 0.007 & 0.357 & 18.709 & 0.004 & 0.240 \\
        2 & 189.47749 & -41.040136 & 19.637 & 0.010 & 0.349 & 18.791 & 0.010 & 0.234 \\
        3 & 189.40136 & -41.048903 & 19.611 & 0.004 & 0.335 & 18.828 & 0.003 & 0.225 \\
        4 & 189.37714 & -40.890137 & 19.998 & 0.003 & 0.359 & 19.251 & 0.016 & 0.241 \\
        \hline
    \end{tabular}
    \label{tab:cen_table}
    \\
    (This table is available in its entirety in a machine-readable form in the online journal. A portion is shown here for guidance regarding its form and content.)
\end{table*}

\begin{table*}[t]
\caption{Eri~IV Photometry in the DELVE Photometric System}
    \begin{tabular}{ccccccccc}
    \hline \hline
        Star No. & $\alpha$ & $\delta$ & $g$ & $\delta g$ & $A_{g}$ & $r$ & $\delta r$ & $A_{r}$ \\
         & (deg J2000.0) & (deg J2000.0) & (mag) & (mag) & (mag) & (mag) & (mag) & (mag) \\
        \hline
        0 & 76.422511 & -9.366627 & 18.784 & 0.001 & 0.326 & 18.055 & 0.003 & 0.234 \\
        1 & 76.558057 & -9.494939 & 18.961 & 0.002 & 0.326 & 18.272 & 0.003 & 0.233 \\
        2 & 76.535325 & -9.493604 & 19.138 & 0.004 & 0.332 & 18.461 & 0.003 & 0.237 \\
        3 & 76.295080 & -9.380901 & 19.561 & 0.002 & 0.339 & 18.911 & 0.004 & 0.243 \\
        4 & 76.341677 & -9.361639 & 19.668 & 0.003 & 0.325 & 19.014 & 0.002 & 0.233 \\
        \hline
    \end{tabular}
    \label{tab:eri_table}
    \\
    (This table is available in its entirety in a machine-readable form in the online journal. A portion is shown here for guidance regarding its form and content.)    
\end{table*}

\section{Analysis} \label{sec:analysis}
\subsection{Color-Magnitude Diagrams} \label{subsec:cmd}

\begin{figure*}[t]
    \centering
    \includegraphics[width=0.24\textwidth]{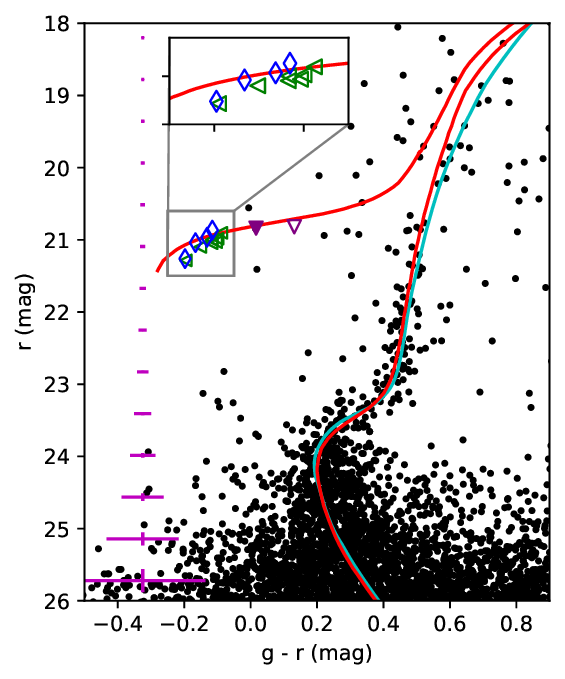}
    \includegraphics[width=0.24\textwidth]{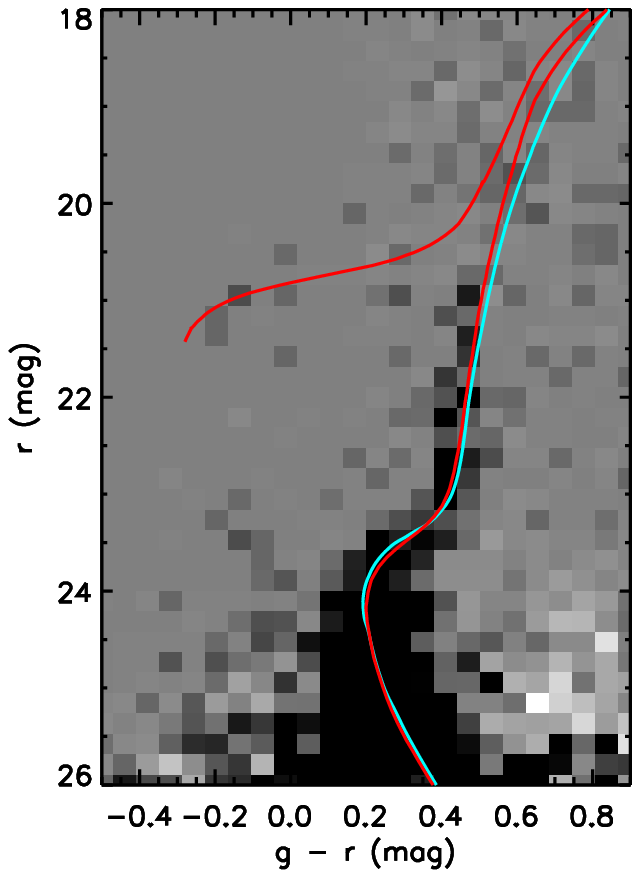}
    \includegraphics[width=0.24\textwidth]{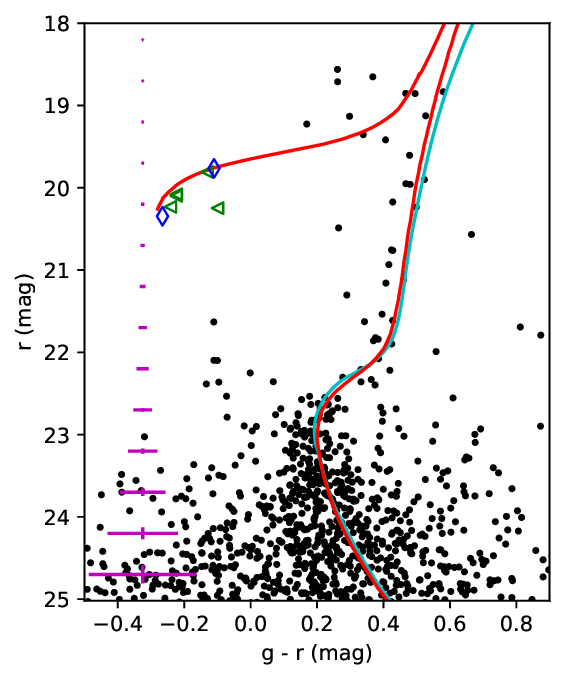}
    \includegraphics[width=0.24\textwidth]{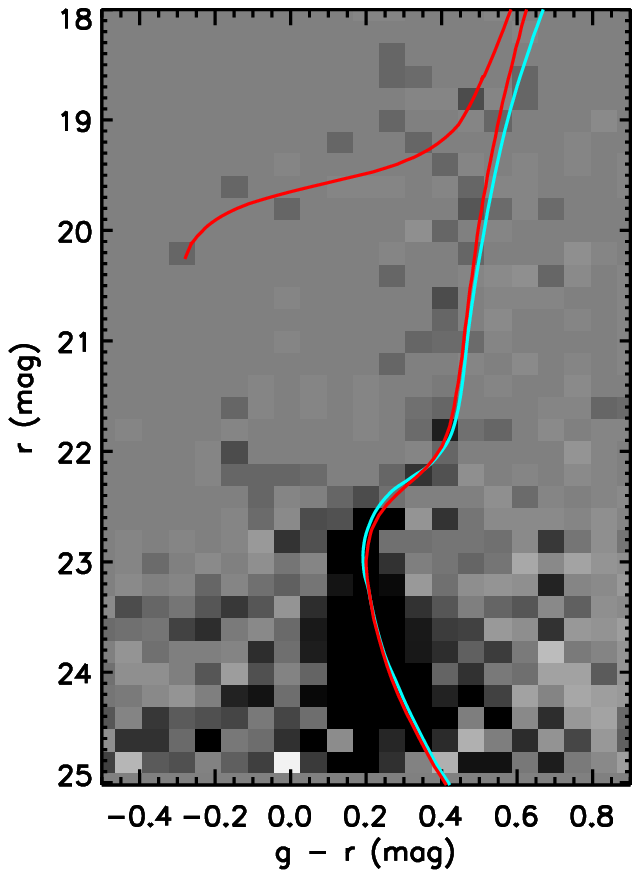}
    
    \hfill\parbox[t]{0.24\textwidth}{(a) Cen I}\hfill
    \parbox[t]{0.24\textwidth}{(b) Eri IV}\hfill
    
    \caption{CMD and Hess diagrams for Cen~I (a) and Eri~IV (b). We only include stars within the half-light radius for a given UFD (Section~\ref{subsec:struct}). Candidate HB stars within the half-light radius are shown as blue diamonds, candidate HB stars within the FoV are green triangles, and RR Lyrae stars from \citet{Martinez-Vazquez2021} are shown as purple triangles for Cen~I only (the filled triangle is within the half-light radius, the other is within the FoV). Magenta error bars show the mean color and magnitude errors in the CMDs. The red lines are PARSEC isochrones with a 13.0 Gyr stellar population and metallicity $\text{[Fe/H]}=-2.2$. The cyan lines are 13.0 Gyr DSEP isochrones with metallicity $\text{[Fe/H]}=-2.4$ and alpha enhancement $[\alpha/\text{Fe}]=0.4$ shown for comparison purposes. The isochrones are shifted to the distance modulus derived in Section~\ref{subsec:dist}: $m-M=\cenDM$ for Cen~I and $m-M=\eriDM$ for Eri~IV.}
    \label{fig:cmds}
\end{figure*}

The left panels of Figure \ref{fig:cmds} (a) and (b) show the color-magnitude diagrams (CMDs) for Cen~I and Eri~IV which include stars within one half-light radius (as determined in Section~\ref{subsec:struct}, see Table \ref{table:struct}). 
Plotted in red are 13.0 Gyr, $\text{[Fe/H]}=-2.2$ PARSEC isochrones \citep{Bressen2012}. 
This corresponds to the best-fit isochrone as determined in Section~\ref{subsec:dist}. 
The magenta error bars show the mean photometric errors in color and magnitude as determined by the artificial star tests (see Section~\ref{sec:data}). These errors are plotted at an arbitrary color for convenience. 
Open blue diamonds are potential horizontal branch (HB) stars within the half-light radius and green triangles are potential HB stars within the FoV: a total of eleven in Cen~I and seven in Eri~IV. 
Newly discovered HB candidates are presented in Tables \ref{table:gaia_cen} and \ref{table:gaia_eri}.
In the CMD for Cen~I we show two RR Lyrae stars from \citet{Martinez-Vazquez2021} as purple triangles; the third star from that study lies outside our FoV. 
The filled triangle lies within our half-light radius, while the other RR Lyrae star is within the FoV.

The right panels of Figure \ref{fig:cmds} (a) and (b) show background-subtracted binned Hess diagrams. Hess diagrams highlight the number density of stars in regions of the CMD. The background is derived from a region beyond a 12\arcmin\xspace radius; this radius is well outside the main body of Cen~I and Eri~IV. The isochrones of the Hess diagrams are the same PARSEC isochrone on the CMDs.

\subsection{Distance} \label{subsec:dist}
We use a CMD-fitting technique to determine the distance modulus, similar to the method described in \citet{Burcin2018}. 
This methodology considers main sequence (MS), red giant branch (RGB), and HB stars; we compare the UFD CMDs with empirical globular cluster (GC) fiducials and theoretical isochrones. 
MS and RGB stars are handled separately from HB stars when deriving the distance modulus. 
For the MS and RGB stars of Cen~I we consider two metal-poor PARSEC isochrones: $\text{[Fe/H]}=-2.2$ and $-2.0$ ($\text{[Fe/H]}=-2.2$ is the lowest metallicity PARSEC isochrone available).
We apply the same PARSEC isochrones to the Eri~IV data.
The stellar ages (13.0 Gyr) of the isochrones were chosen to be consistent with the discovery papers. 
We shift the distance modulus ($m-M$) applied to the isochrone in 0.025~mag intervals over 2 magnitudes. 
With each shift, we count the number of stars that are consistent with the isochrone when taking into account the photometric uncertainties of the data. 
When the photometric errors are $<0.1$~mag, we inflate the uncertainty to 0.1~mag. 

For Cen~I, we select all stars with $r\leq24$~mag that are within one half-light radius ($r_{h}=\cenRH\arcmin$) of its center.  
We shift the distance modulus from 19.5 mag to 21.5 mag; this is consistent with the distance modulus reported by the discovery paper. 
We also run this procedure over background stars selected from an equal area background region offset $\gtrsim11$\arcmin\xspace from the center of Cen~I. 
The best-fit distance modulus is when the isochrone fit yields the maximum number of stars after accounting for background contamination: $m-M=20.45$ for $\text{[Fe/H]}=-2.2$ (118 stars), and $m-M=20.30$ for $\text{[Fe/H]}=-2.0$ (122 stars).
Likewise, for Eri~IV we select stars with $r\leq23$~mag and within one half-light radius ($r_h=\eriRH\arcmin$) of its center. 
The distance modulus is shifted from 18.5~mag to 20.5~mag, and we account for background stars similarly to Cen~I. 
The maximizing distance moduli for Eri~IV are $m-M=19.23$ for $\text{[Fe/H]}=-2.2$ (37 stars), and $m-M=19.21$ for $\text{[Fe/H]}=-2.0$ (38 stars).
We determine the uncertainties on each fit using a 100 iteration bootstrap resampling analysis.


We also derive a distance modulus for both dwarfs using their potential HB stars within the FoV. This is 11 stars for Cen~I and 7 stars for Eri~IV (see Figure~\ref{fig:cmds}). 
The number of potential HB stars in Cen~I and Eri~IV is comparable to other well-studied systems \citep[e.g.][]{Sand2010, Koposov2015, Munoz2018, Homma2019}.
We fit the M92 globular cluster HB fiducial \citep{Bernard2014} to our HB star candidates by minimizing the sum of squares of the difference between the data and the fiducial. 
We measure the $r$ magnitude offset a given HB star lies from the fiducial, shift the distance modulus of the fiducial by 0.025 mag, and remeasure the offset. 
This process is repeated through two magnitudes. The shift that minimizes the sum of squares is adopted as the distance modulus. 
We opt to use empirical HB fiducial tracks rather than HB isochrones as HB isochrones are historically more difficult to model and have a certain level of uncertainty associated with them \citep[e.g.][]{Pietrinferni2004}. 
We use M92 ($m-M=14.65$ mag, \citealt{Bernard2014}) specifically because it is one of the most ancient, metal-poor, and well-studied globular clusters.

Using this HB distance modulus measurement technique, we find $m-M=20.43$ for Cen~I and $m-M=19.18$ for Eri~IV. 
To determine the uncertainty associated with the HB distance modulus we implement jackknife resampling (i.e., we remove a single HB star from the sample and remeasure the distance modulus). 
This technique accounts for the possibility of interloper stars contaminating our HB sample. 
The standard deviation of the jackknife resampled distance moduli is adopted as the uncertainty for this measurement. 
Jackknife resampling does not change the distance modulus measurement for Cen~I. 
The distance modulus for Eri~IV ranges from $19.15\leq m-M \leq 19.18$ with a standard deviation of $\sim0.01$ after resampling.

To find the total distance modulus errors, we add (in quadrature) the uncertainty derived from the bootstrap analysis applied to MS/RBG stars and the uncertainty from the jackknife-resampled HB stars. The distance modulus we adopt for Cen~I and Eri~IV is the mean value between the two methods (MS/RGB star counting and HB fiducial fitting); these results are shown in Table \ref{table:struct}.

Old, low-metallicity isochrones exhibit minimal variation based on the chosen age and metallicity. 
Separate libraries, such as the Dartmouth Stellar Evolution Database \citep[DSEP;][]{Dotter2008}, closely resemble PARSEC isochrones in this regime.
Using DSEP isochrones ($\text{[Fe/H]}=-2.4, -2.2$ and $[\alpha/\text{Fe]}=0.4$) we derive distances consistent with the PARSEC models within the uncertainties quoted in Table \ref{table:struct}; we show one DSEP isochrone in Figure \ref{fig:cmds} for reference. 
Ultimately, we choose the PARSEC models to remain consistent with the discovery papers.

\subsection{Structural Properties} \label{subsec:struct}

\begin{table}[]
\centering
\caption{Structural Properties of Cen~I and Eri~IV}
\begin{tabular}{l l l}
 \hline \hline
 Parameter & Cen~I & Eri~IV \\ 
 \hline
 
  $\alpha_{\textrm{2000}}$ (deg) & \cenRA $\pm$\cenRAerr & \eriRA$\pm$\eriRAerr \\
   $\delta_{\textrm{2000}}$ (deg) & \cenDEC$\pm$\cenDECerr & \eriDEC$\pm$\eriDECerr \\
  $m-M$ (mag) & \cenDM$\pm$\cenDMerr & \eriDM$\pm$\eriDMerr \\
   D (kpc) & \cenDKPC$\pm$\cenDKPCerr & \eriDKPC$\pm$\eriDKPCerr \\
   $M_{V}$ (mag) & \cenMV$\pm$\cenMVerr & \eriMV$\pm$\eriMVerr \\
   $r_{h}$ (arcmin) & \cenRH$\pm$\cenRHerr & \eriRH$\pm$\eriRHerr \\
   $r_{h}$ (pc) & \cenRHPHYS$\pm$\cenRHPHYSerr & \eriRHPHYS$\pm$\eriRHPHYSerr \\
   $r_{1/2}$ (arcmin) & \cenRGEO$\pm$\cenRGEOerr & \eriRGEO$\pm$\eriRGEOerr \\
   $r_{1/2}$ (pc) & \cenRGEOphys$\pm$\cenRGEOphyserr & \eriRGEOphys$\pm$\eriRGEOphyserr \\
   $\epsilon$ & \cenELLIP$\pm$\cenELLIPerr & \eriELLIP$\pm$\eriELLIPerr \\
   P.A. (deg) & \cenTHETA$\pm$\cenTHETAerr & \eriTHETA$\pm$\eriTHETAerr \\
   \hline 
   $v_{sys}$ (km s$^{-1}$) & 44.9$\pm$0.8 & -31.5$^{+1.3}_{-1.2}$ \\
   $\sigma_{v}$ (km s$^{-1}$) & 4.2$^{+0.6}_{-0.5}$ & 6.1$^{+1.2}_{-0.9}$ \\
   
   [Fe/H] (dex) & $-2.57\pm0.08$ & $-2.87^{+0.08}_{-0.07}$ \\
   $\sigma_{\text{[Fe/H]}}$ (dex) & $0.38^{+0.07}_{-0.05}$ & $0.20\pm0.09$ \\
    $\mu_{\alpha} \cos \delta$ (mas yr$^{-1}$) & \cenPMRA$\pm$\cenPMRAerr & \eriPMRA$\pm$\eriPMRAerr \\
   $\mu_{\delta}$ (mas yr$^{-1}$) & \cenPMDEC$\pm$\cenPMDECerr & \eriPMDEC$\pm$\eriPMDECerr \\
   $r_{peri}$ (kpc) & $32^{+12}_{-8}$ & $43\pm11$ \\
 \hline
\end{tabular}
 \begin{tablenotes}
      \small
      \item  $\alpha_{\textrm{2000}}$: the Right Ascension (J2000.0). $\delta_{\textrm{2000}}$: the Declination (J2000.0). $m-M$: the distance modulus. $D$: the distance of the galaxy in kpc. $M_{V}$: the absolute V-band magnitude. $r_{h}$: the elliptical half-light radius along the semi-major axis. $r_{1/2}$: the geometric mean of the half-light radius. $\epsilon$: ellipticity which is defined as $\epsilon=1-b/a$, where $b$ is the semiminor axis and $a$ is the semimajor axis. P.A.: Position angle. $v_{sys}$: systemic radial velocity in the heliocentric frame. $\sigma_{v}$: velocity dispersion. [Fe/H]: mean metallicity.  $\sigma_{\text{[Fe/H]}}$: metallicity dispersion. $\mu_{\alpha} \cos \delta$: Systemic proper motion in R.A. $\mu_{\delta}$: Systemic proper motion in Dec. $r_{peri}$: orbital pericenter. Values $\alpha_{\text{2000}}$ through the position angle (P.A.) are calculated in this work; values $v_{sys}$ through $r_{peri}$ are from \citet{Heiger2023}.
    \end{tablenotes}
\label{table:struct}
\end{table}

To determine the structural properties of Cen~I and Eri~IV, we use a maximum likelihood analysis described in \citet{Sand2009} \citep[see also][]{Martin2008}. 
In short, the routine fits an exponential profile to the 2D distribution of stars associated with each UFD. 
We select stars consistent within our 90\% completeness measurement and a $\text{[Fe/H]}=-2.2$ PARSEC isochrone in color-magnitude space after accounting for photometric uncertainties. 
This is 812 stars for Cen~I and 243 stars for Eri~IV.
The structural parameters from the discovery papers \citep{Mau2020, Cerny2021} are adopted as the input parameters for the initial analysis. 
The data are bootstrap resampled 1000 times; the structural parameters are recalculated for each such resampling which determines the uncertainty on the measurement. 
The resulting structural parameters are shown in Table~\ref{table:struct}.

\begin{figure*}
    \includegraphics[width=0.5\textwidth, height=0.5\textwidth]{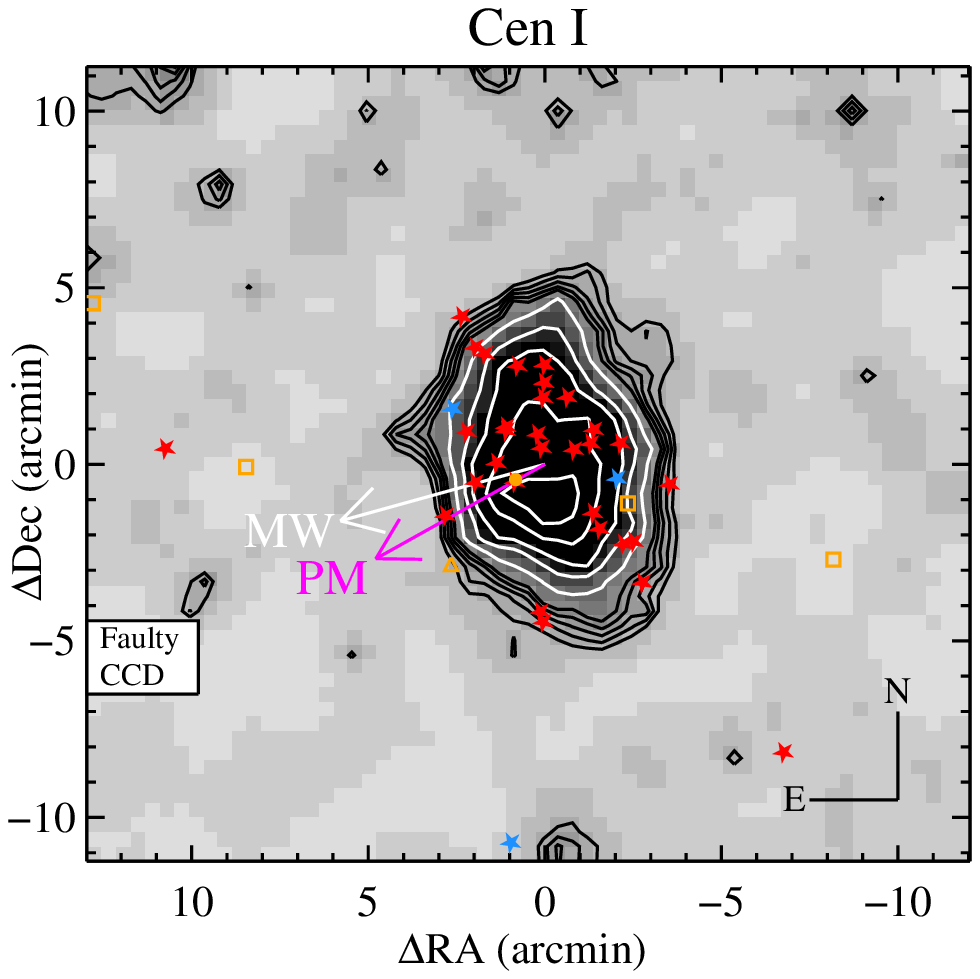}
    \includegraphics[width=0.5\textwidth, height=0.5\textwidth]{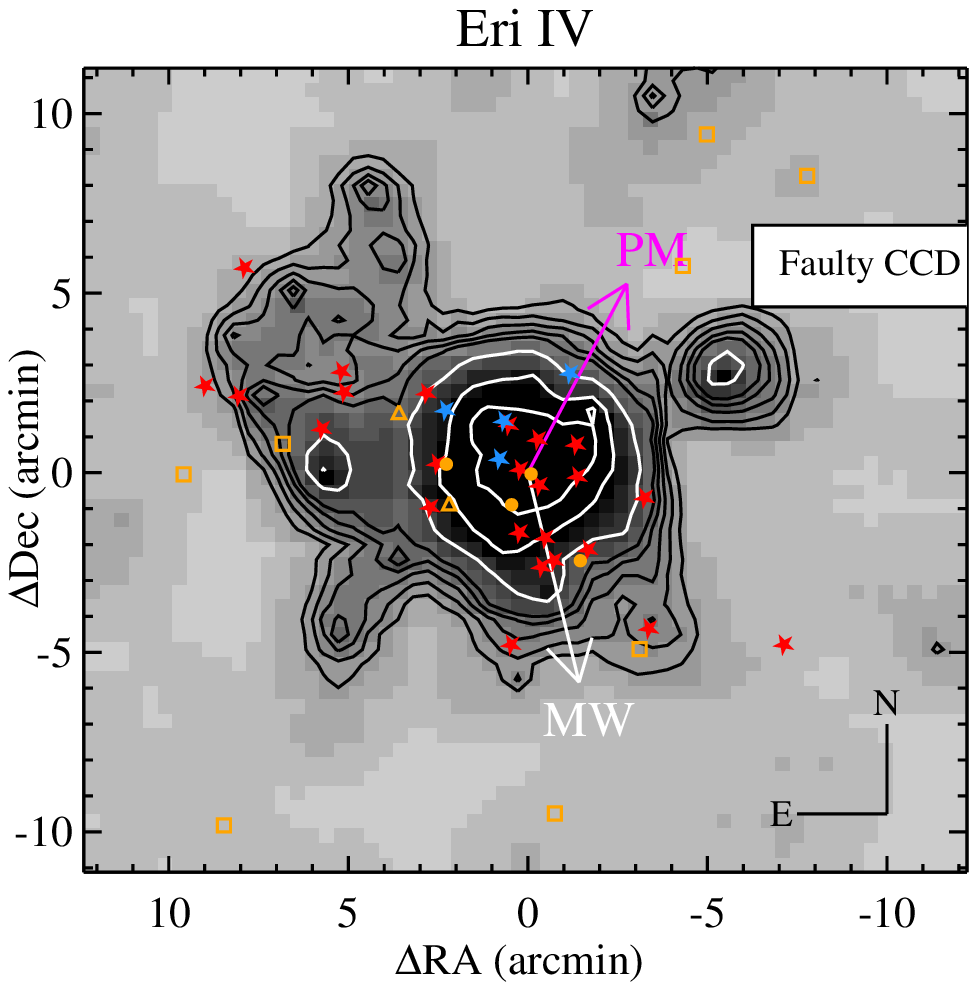}
    \caption{The smoothed matched-filter maps for Cen~I (left) and Eri~IV (right) where the image center corresponds to the RA and DEC reported in Table \ref{table:struct}; north is up and east is to the left.
    Overplotted are the contour levels above the root mean square of the background for each UFD which corresponds to $3\sigma$, $4\sigma$, $5\sigma$, $6\sigma$, $7\sigma$, $10\sigma$, $15\sigma$, $20\sigma$, $30\sigma$, $40\sigma$ for Cen~I and $3\sigma$, $4\sigma$, $5\sigma$, $6\sigma$, $7\sigma$, $10\sigma$, $15\sigma$, $20\sigma$ for Eri~IV. The white arrow denotes the direction to the Galactic center and the magenta arrow marks the solar-reflex-corrected proper motion. The blue and red stars represent the location of member stars observed by \citet{Heiger2023}, where blue are HB stars and red are RGB stars. New potential members (see Section \ref{subsec:proper_motion}) are shown in orange: circles are found using the conservative method, squares using the flexible method, and triangles are recovered with both methods. Note that faulty CCDs are shown in the lower left hand side for Cen~I, and the upper right hand side for Eri~IV.}
    \label{fig:ext_struct}
\end{figure*}

\subsection{Absolute Magnitude} \label{subsec:abs_mag}

We derive the absolute magnitudes for Cen~I and Eri~IV by using the same procedure as in \citet{Burcin2018}, as was first described in \citet{Martin2008}. We construct a densely populated (N$\sim$45,000) artificial CMD where stars are consistent with our completeness and photometric uncertainties. 
The luminosity functions are generated using the PARSEC database with $\text{[Fe/H]}=-2.2$, and we assume a Salpeter initial mass function \citep[IMF;][]{Salpeter1955}.
We randomly select $N$ number of stars from our artificial CMD (over the same magnitude range as was used to derive the structural properties), where $N$ was derived from the exponential profile fits (Section \ref{subsec:struct}). We calculate the total luminosity by adding the fluxes of these randomly selected stars and estimating the flux of the faint, unresolved component of the galaxy using the adopted luminosity function. We perform 1000 realizations in this way, and take the mean as our absolute magnitude and its standard deviation as our uncertainty. We account for variation in the distance modulus and number of stars by allowing either value to vary by their associated errors; this process is repeated 100 separate times. The errors are added in quadrature to produce the total uncertainty on our absolute magnitude measurement. 
We account for HB stars by adding the fluxes of the HB candidates within our FoV. This yields an absolute magnitude of  $M_{V}=\cenMV\pm\cenMVerr$ mag for Cen~I and $M_{V}=\eriMV\pm\eriMVerr$ mag for Eri~IV (see Table \ref{table:struct}). For each system, the impact of HB candidates on the absolute magnitude falls within the stated uncertainty.

\subsection{Extended Structure Search} \label{subsec:ext_struct}

We investigate whether our targets have undergone tidal disruption, which might manifest in the form of extended structures and/or stellar streams. This analysis holds particular significance as the discovery data suggested the presence of extended structures; however, subsequent spectroscopic observations \citep{Heiger2023} disfavor tidal disruption in Cen~I or Eri~IV. We use a matched-filter algorithm, which maximizes the signal-to-noise ratio of data against the background. The concept of applying matched-filter algorithms to study tidal disruptions was introduced by \citet{Rockosi2002} and initially applied to GCs. This approach has since been adopted to investigate newly identified UFDs for signs of extended structures \citep[e.g. ][]{Sand2012, Burcin2018}.

To apply the matched-filter technique to our data, we use the same simulated stars as used in Section~\ref{subsec:abs_mag} as the signal CMD (i.e., stars simulated to match an old, metal-poor PARSEC isochrone, accounting for photometric errors and magnitudes brighter than the 90\% completeness threshold). 
Simulated data is preferable to the observed data which can be sparsely populated and contaminated with background/foreground objects \citep{Sand2012}. 
For the background CMD, we use the real stars well outside the half-light radius of both UFDs. 
The matched-filter stellar density maps are shown in Figure~\ref{fig:ext_struct}; both maps have been spatially binned to a pixel size of 25\arcsec\xspace and smoothed with a Gaussian width 1.0 times the pixel size. 
The background and variance of the smoothed maps are determined with IDL's MMM routine. The main body of each satellite is clearly visible in each map. 
The white arrows in Figure \ref{fig:ext_struct} represent the direction to the Galactic center, and the magenta arrows denote the solar-reflex-corrected proper motions from \citet{Heiger2023}. 
The spectroscopic member stars observed by \citet{Heiger2023} are shown in red (RGB) and blue (HB), and the potential member stars discussed in Section \ref{subsec:proper_motion} are shown in orange.

Both Cen~I and Eri~IV present as clear overdensities against the background. Cen~I exhibits a well-defined morphology lacking any clear extended features. For Eri~IV, there appears to be a significant feature positioned to the northeast of the main body. This feature shows up as a $3\sigma, 4\sigma, 5\sigma, 6\sigma,$ and $7\sigma$ overdense region depending on the location. We use the statistical tests outlined in \citet{Walsh2008} and \citet{Sand2010} to assess whether this feature results from small number of statistics or background features. 
We bootstrap resample Eri~IV's entire photometric catalog, and we resample the stars which are consistent with the isochrone and uncertainties. 
For each such resampling we update the smoothed maps and inspect regions of particular interest.
These tests recover the extended feature for all of the resamplings. 
The ``nugget" situated below the faulty CCD varies in significance during the resamplings and goes away in a third of the tests.
Additionally, we visually inspect these regions in the images and find no artifacts or galaxy clusters which may masquerade as a stellar overdensity.
We keep this in mind when considering the overall morphology of Eri~IV, and discuss its implications in more depth in Section \ref{sec:discussion}.

\subsection{Potential Member Stars in Gaia} \label{subsec:proper_motion}
We use our deep photometry and the precise astrometry of \textit{Gaia} Early Data Release 3 (EDR3; \citealt{Gaia2016}, \citealt{Gaia2021}) to search for potential new members stars using two separate methods. 
The first method cross matches \textit{Gaia} objects with our photometry after applying a CMD filter described in \citet{Pace2019} and \citet{Pace2022}.
A Gaussian mixture model composed of a MW background and a satellite component is applied to the proper motions and spatial positions of these stars.
Similar to \citet{Pace2022}, we assume Gaussian priors for the structural parameters derived in Section \ref{subsec:struct}.
This conservative method weights the probability of a star's membership based on its distance from the center of the UFD. We find two probable member stars in Cen~I and six in Eri~IV.

We also explore the entire FoV by cross-matching stars that agree with our CMD selection used in Section \ref{subsec:struct} with \textit{Gaia} EDR3. We keep stars that satisfy the following criteria: 
{\fontfamily{cmtt}\selectfont ruwe}~$<1.4$, {\fontfamily{cmtt}\selectfont astrometric\_excess\_noise}~$<2$, and stars which are consistent with zero parallax ($\varpi - 3\sigma_{\varpi} > 0$).
Furthermore, we exclude stars which are inconsistent with the proper motions of the stars presented in \citet{Heiger2023}. 
This flexible selection method yields four additional stars in Cen~I and eight additional stars in Eri~IV. 
The potential new member stars in Cen~I and Eri~IV are listed in Tables \ref{table:gaia_cen} and \ref{table:gaia_eri}, respectively. They are ideal targets for a future spectroscopic study. Their spatial positions are shown in Figure \ref{fig:ext_struct} in orange.
Two of Eri~IV's newly identified potential member stars are situated near the extended feature.


\begin{table*}[b]
\caption{Newly identified potential member stars in Cen~I. Method 1 uses the conservative selection and method 2 uses the flexible selection described in Section \ref{subsec:proper_motion}. Entries with no \textit{Gaia} ID are potential HB stars used in the distance determination (Section \ref{subsec:dist}), but do not have \textit{Gaia} observations.}
\centering
    \begin{tabular}{ccccccccc}
    \hline \hline
        Gaia ID & RA & Decl. & g & r & $\mu_{\alpha}\cos\delta$ & $\mu_{\delta}$ & Type & Method \\
         & (deg) & (deg) & (mag) & (mag) & (mas yr$^{-1}$) & (mas yr$^{-1}$) \\
        \hline
        6146244710500639104 & 189.454650 & -40.949242 & 20.64 & 20.10 & -1.474 $\pm$ 0.632 &  0.972 $\pm$ 0.589 &  RBG &   2 \\
                  - & 189.536440 & -40.870731 & 20.88 & 21.00 &                  - &                  - &   HB$^{\text{*}}$ &   - \\
                  - & 189.547430 & -40.910299 & 20.93 & 21.03 &                  - &                  - &   HB$^{\text{*}}$ &   - \\
                  - & 189.550540 & -40.957694 & 20.89 & 20.99 &                  - &                  - &   HB$^{\text{*}}$ &   - \\
        6146244366903299328 & 189.551620 & -40.922743 & 18.86 & 18.10 & -2.851 $\pm$ 0.165 & -1.761 $\pm$ 0.144 &  RGB &   2 \\
                  - & 189.567240 & -40.883106 & 20.75 & 20.87 &                  - &                  - &   HB$^{\text{*}}$ &   - \\
                  - & 189.587720 & -40.935858 & 21.06 & 21.26 &                  - &                  - &   HB$^{\text{*}}$ &   - \\
                  - & 189.596760 & -40.900816 & 20.87 & 21.04 &                  - &                  - &   HB$^{\text{*}}$ &   - \\
                  - & 189.603500 & -40.918177 & 20.83 & 20.96 &                  - &                  - &   HB$^{\text{*}}$ &   - \\
        6146234097636409344 & 189.604605 & -40.911450 & 20.38 & 19.72 &  0.171 $\pm$ 0.489 & -0.479 $\pm$ 0.414 &  RGB &   1 \\
                  - & 189.617670 & -40.960118 & 20.95 & 21.10 &                  - &                  - &   HB$^{\text{*}}$ &   - \\
                  - & 189.626020 & -40.850364 & 20.93 & 21.05 &                  - &                  - &   HB$^{\text{*}}$ &   - \\
                  - & 189.632750 & -40.839181 & 21.09 & 21.28 &                  - &                  - &   HB$^{\text{*}}$ &   - \\
        6146233754039010048 & 189.635114 & -40.951360 & 19.55 & 18.89 & -0.160 $\pm$ 0.321 & -0.469 $\pm$ 0.250 &  RBG &  1, 2 \\
                  - & 189.643710 & -40.905551 & 20.81 & 20.90 &                  - &                  - &   HB$^{\text{*}}$ &   - \\
        6146233406146774784 & 189.731830 & -40.905546 & 19.33 & 18.66 & -1.095 $\pm$ 0.234 & -0.890 $\pm$ 0.224 &  RGB &   2 \\
        6146257457963749248 & 189.805530 & -40.828194 & 20.57 & 20.02 & -0.916 $\pm$ 0.703 & -1.049 $\pm$ 0.651 &  RGB &   2 \\
        \hline
        \end{tabular}
        \label{table:gaia_cen}
        \begin{tablenotes}
            \item $^{\text{*}}$ Denotes star used in distance determination.
        \end{tablenotes}
\end{table*}

\begin{table*}[b]
\caption{Newly identified potential member stars in Eri~IV. Method 1 uses the conservative selection and method 2 uses the flexible selection described in Section \ref{subsec:proper_motion}.}
\centering
    \begin{tabular}{ccccccccc}
    \hline \hline
        Gaia ID & RA & Decl. & g & r & $\mu_{\alpha}\cos\delta$ & $\mu_{\delta}$ & Type & Method \\
         & (deg) & (deg) & (mag) & (mag) & (mas yr$^{-1}$) & (mas yr$^{-1}$) \\
        \hline
        3182741431156735488 & 76.295080 & -9.380901 & 19.22 & 18.67 &  1.305 $\pm$ 0.246 &  0.300 $\pm$ 0.206 &    RGB &    2 \\
        3182741671674916608 & 76.341677 & -9.361639 & 19.34 & 18.78 &  1.749 $\pm$ 0.213 &  0.497 $\pm$ 0.196 &    RGB &    2 \\
        3182738063902340736 & 76.352797 & -9.422692 & 19.59 & 19.05 &  0.065 $\pm$ 0.280 &  0.053 $\pm$ 0.240 &    RGB &    2 \\
        3182721639948162176 & 76.372800 & -9.600543 & 19.68 & 19.81 &  0.768 $\pm$ 0.468 & -0.609 $\pm$ 0.381 &     HB$^{\text{*}}$ &    2 \\
        3182722911258501888 & 76.400357 & -9.559472 & 19.94 & 19.70 &  0.066 $\pm$ 0.399 &  0.037 $\pm$ 0.353 & HB/RRL &    1 \\
        3182718135254812672 & 76.412174 & -9.676816 & 19.47 & 18.92 &  0.749 $\pm$ 0.271 & -0.038 $\pm$ 0.238 &    RGB &    2 \\
        3182724629244570496 & 76.423206 & -9.519234 & 19.35 & 18.94 &  0.121 $\pm$ 0.257 & -0.031 $\pm$ 0.236 &    RGB &    1 \\
        3182724457445870848 & 76.432246 & -9.533512 & 19.87$^{a}$ & 20.07$^{a}$ &  0.859 $\pm$ 0.525 & -0.564 $\pm$ 0.500 &     HB &    1 \\
        3182723843267791616 & 76.461253 & -9.532902 & 19.32 & 18.85 &  0.282 $\pm$ 0.237 & -0.036 $\pm$ 0.212 &    RGB &   1, 2 \\
        3182724972841956608 & 76.462523 & -9.514698 & 20.60 & 20.17 &  0.302 $\pm$ 0.764 &  0.062 $\pm$ 0.704 &    RGB &    1 \\
        3182725114578785792 & 76.484535 & -9.490527 & 19.87 & 20.10 &  0.908 $\pm$ 0.526 & -0.219 $\pm$ 0.506 &     HB$^{\text{*}}$ &   1, 2 \\
        3182747581549809024 & 76.538458 & -9.505236 & 19.85 & 20.08 &  0.278 $\pm$ 0.549 &  0.345 $\pm$ 0.503 &     HB$^{\text{*}}$ &    2 \\
        3182670443938116864 & 76.565750 & -9.682317 & 20.87 & 20.41 &  0.850 $\pm$ 0.791 &  0.655 $\pm$ 0.850 &    RGB &    2 \\
        3182744523533084416 & 76.584477 & -9.519372 & 20.15 & 20.25 & -0.303 $\pm$ 0.544 & -0.279 $\pm$ 0.445 &     HB$^{\text{*}}$ &    2 \\
        \hline
        \end{tabular}
        \label{table:gaia_eri}
        \begin{tablenotes}
            \item $^{a}$ This star falls in a chip gap and has unreliable photometry, so we report the DELVE DR3 photometry.
            \item $^{\text{*}}$ Denotes star used in distance determination.
        \end{tablenotes}
\end{table*}

\section{Discussion} \label{sec:discussion}

\begin{figure*}[]
    \centering
    \includegraphics[width=0.75\textwidth]{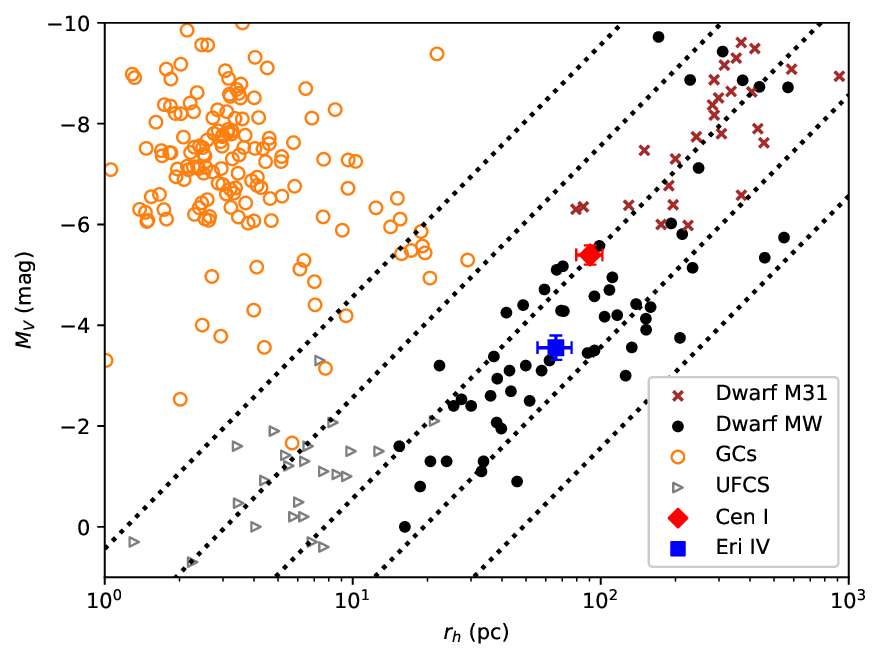}
    \caption{Half-light radius along the semi-major axis ($r_{h}$) against the absolute magnitude ($M_{V}$) for a number of dwarf galaxies/globular clusters. Cen~I and Eri~IV are the red diamond and blue square, respectively. Brown x's are M31 dwarf galaxies \citep[e.g.][]{Tollerud2012, Martin2016, Rhode2023}, black circles are MW dwarf galaxies, orange open circles are GCs from \citet{Harris2010}, and gray triangles are ultra-faint compact satellites (UFCS). The dotted black lines denote a constant surface brightness. Data in this figure were taken from the Local Volume Database \citep{Pace2024}.}
    \label{fig:size_luminosity}
\end{figure*}

Figure~\ref{fig:size_luminosity} shows Cen~I (red diamond) and Eri~IV (blue square) in the size-luminosity plane, relative to Local Group dwarf galaxies and GCs. 
Black circles are the most up-to-date confirmed/candidate MW dwarf satellites \citep[see:][]{McConnachie2012, DW2015, Kim2015, Koposov2015, Martin2015, Crnojevic2016, DW2016, Torrealba2016a, Torrealba2016b, Carlin2017, Choi2018, Homma2018, Koposov2018, Munoz2018, Burcin2018, Torrealba2018, Homma2019, Wang2019, Mau2020, Moskowitz2020, Simon2020, Cantu2021, Cerny2021, Ji2021, Richstein2022, Cerny2023a, Cerny2023b, Homma2023, Smith2023}. The data in Figure~\ref{fig:size_luminosity} were taken from the  Local Volume Database \citep{Pace2024}. Both Cen~I and Eri~IV are consistent with the population of known dwarf galaxies. We discuss each satellite’s derived properties in detail in the following subsections.

\subsection{Cen~I} \label{subsec:disc_cen1}

The deep imaging of Cen~I shows a well-populated MS, an easily identifiable RGB, eleven potential HB candidates within the FoV (four within one $r_{h}$), and a handful of blue stragglers (Figure~\ref{fig:cmds}, left). 
The stellar population is consistent with an age of $\tau\approx13.0$ Gyr and a metallicity of $\text{[Fe/H]}\sim-2.2$. We find the distance to Cen~I is $D=\cenDKPC\pm\cenDKPCerr$~kpc ($m-M=\cenDM\pm\cenDMerr$~mag); this is in good agreement with the results reported in \citet{Mau2020} ($D=116.3^{+1.6}_{-0.6}$~kpc, $m-M=20.33^{+0.03}_{-0.01}\pm0.1$~mag). 
Our measurement also agrees well with \citet{Martinez-Vazquez2021} who derived the distance using three RR Lyrae stars. They found $D=117.7\pm0.1$~kpc, $m-M=20.354\pm0.002$~mag with systematic errors of 4~kpc and 0.07~mag.
We show two of the RR Lyrae stars in Figure~\ref{fig:cmds}, the third star was outside our FoV.
Three RR Lyrae stars is consistent with the expected number ($1-12$ stars) for UFDs of similar magnitudes \citep{Martinez-Vazquez2019, Martinez-Vazquez2021}.

We robustly constrain the structural parameters (i.e. RA, DEC, $r_{h}$, $\epsilon$, P.A.) of Cen~I using a maximum likelihood analysis (see Section~\ref{subsec:struct}); these results are shown in Table \ref{table:struct}. Our structural parameters agree very well with the values reported in the discovery data with decreased uncertainties on most of the measurements. 
Cen~I occupies a space on the size-luminosity plane consistent with other confirmed UFDs (Figure~\ref{fig:size_luminosity}). 
Our work further supports the UFD nature of Cen~I and strengthens the findings of \citet{Mau2020} and \citet{Heiger2023}. 

Dynamical interactions between satellites and their host galaxies, such as tidal stripping, serve as feedback mechanisms which reduce the central mass of satellites. 
Mass-to-light (M/L) ratios are calculated under the assumption of dynamical equilibrium. 
If a system is tidally disrupting, then dark matter estimates (via velocity dispersion) would be inaccurate. 
Therefore, it is crucial to understand the dynamical state and morphologies of UFDs. 
Cen~I has a well-defined structure and exists as a clear overdensity in Figure~\ref{fig:ext_struct}. 
\citet{Mau2020} identified a potential tidal feature in the discovery data situated $\sim10$\arcmin\xspace west of Cen~I's center; however, we do not find any evidence for the existence of this feature in our deep, wide-field data. 
The feature suggested in the discovery paper is potentially due to background contamination in shallow data. \citet{Heiger2023} did not observe this region spectroscopically, as none of the high-probability member stars reside in this region. 
The spectroscopic analysis revealed a velocity gradient consistent with zero across Cen~I's semi-major axis (tidally disrupting systems are expected to have a large velocity gradient in the direction of its orbit, e.g. \citealt{Ji2021}). Therefore, \citet{Heiger2023} conclude that Cen~I is not tidally disrupting based on the spectroscopic evidence.
Due to the lack of any overdensity in the proposed region of disruption, our findings agree with \citet{Heiger2023} and we conclude that Cen~I does not have a tidal feature associated with it. 


\subsection{Eri~IV} \label{subsec:disc_eri4}  

The CMD of Eri~IV shows a MS with a fair amount of color scatter (due to photometric uncertainties), but there exists a clear MS-turnoff. 
The RGB is less populated compared to Cen~I, which aligns with our findings that Cen~I is brighter. 
We find seven potential HB stars within the FoV (two within one $r_{h}$) and a number of potential blue stragglers (Figure \ref{fig:cmds}). 
We find one potential RR Lyrae star (Table \ref{table:gaia_eri}) which appears in the \textit{Gaia} DR3 variability catalog \citep{Eyer2023}, but is classified as an eclipsing binary. A system with the luminosity of Eri~IV is expected to have few, if any, RR Lyrae stars \citep{Martinez-Vazquez2019}.
The stellar population of Eri~IV is consistent with an age of $\tau\approx13.0$~Gyr and a metallicity of $\text{[Fe/H]}\lesssim-2.2$. We find Eri~IV to be marginally closer than initially reported, but consistent within the joint $1\sigma$ uncertainty: $D=\eriDKPC\pm\eriDKPCerr$~kpc (this work), compared to $D=76.7^{+4.0}_{-6.1}$~kpc from \citet{Cerny2021}.

Our structural parameters mostly agree with those reported with the discovery data, i.e., $\alpha_{\text{2000}}$, $\delta_{\text{2000}}$, $r_{h}$, and the position angle. Our data reveal a less elongated shape for Eri~IV: we find $\epsilon=\eriELLIP\pm\eriELLIPerr$ and \citet{Cerny2021} found $\epsilon=0.54^{+0.10}_{-0.14}$. We derive an absolute magnitude of $M_{V}=\eriMV\pm\eriMVerr$~mag which makes Eri~IV fainter than what was previously thought ($M_{V}=-4.7\pm0.2$ mag, \citealt{Cerny2021}). However, had we used the distance modulus from the discovery data ($m-M=19.42^{+0.01}_{-0.08}\pm0.1$), we \textit{would have} calculated $M_{V}=-3.81\pm0.27$~mag which agrees with our measurement.

Eri~IV displays a well-defined central body, a ``nugget" situated below the faulty CCD, and an extended feature located to the northeast of the main system (Figure~\ref{fig:ext_struct}, right). The nugget varies in significance depending on the smoothing/binning parameters, and it goes away in approximately one-third of our resampling tests. 
Additionally, none of the spectroscopic member stars or the potential member stars are aligned with this feature.
For these reasons, and its proximity to the faulty CCD, we cannot confidently claim this feature is real.
The extended feature is situated in the same area as the overdensity described in the discovery paper \citep[][their Figure 1]{Cerny2021}. The authors advised careful interpretation of this feature due to the limited number of observed stars. \citet{Heiger2023} spectroscopically observed stars near the main body ($N=22$) and the extended feature ($N=6$) of Eri~IV. They find a velocity gradient consistent with zero along the semi-major axis. They estimate the tidal radius at its pericenter is $r_{t}=498\pm128$~pc which yields $r_{t}/r_{h}\approx 6.6\pm 0.3$. A system undergoing tidal disruption is thought to have a tidal radius that is approximately equal to its half-light radius \citep{Pace2022}, and the tidal radius is often approximated as the Jacobi radius \citep{Binney2008}. These reasons lead \citet{Heiger2023} to conclude that Eri~IV is likely not tidally disrupting. Our updated half-light radius increases the ratio to $r_{t}/r_{h}\approx7.6$. It is worth noting that some cosmological simulations \citep[e.g. FIRE;][]{Wetzel2016} find that satellites disrupt well outside the $r_{t}/r_{h}\approx1$ limit, and the Jacobi radius/tidal radius approximation does not always hold \citep{Shipp2023}.

Our data suggest that the extension northeast of Eri~IV's center is significant, and could \textit{potentially} be the result of tides.
This feature persists regardless of the smoothing or binning parameters used to generate the matched-filter map. 
We also note that the spectroscopically confirmed member stars observed by \citet{Heiger2023} closely align with this feature, along with two of the potential new members. 
The extension persists with similar significance when we bootstrap resample Eri~IV's stars (see Section~\ref{subsec:ext_struct}). 
This persistence confirms that this region is not the result of background noise or small number statistics. 
Given that the high significance of this feature is consistently seen across different analysis techniques/parameters and is distinguishable from background noise, it is possible that Eri~IV may have experienced tidal interactions leading to the extended structure visible in Figure \ref{fig:ext_struct} (right). 

The spectroscopic evidence does not favor a tidal stripping scenario. 
The potential tidal debris is misaligned with the predicted orbit of Eri~IV, and \citet{Heiger2023} show that including/excluding the Large Magellanic Cloud does not significantly alter this prediction.
Furthermore, \citet{Heiger2023} show that Eri~IV is moving towards its pericenter ($r_{\text{peri}}=43\pm11$ kpc) having been at its apocenter ($r_{\text{apo}}=135^{+24}_{-13}$ kpc) approximately 1 Gyr ago. It is unlikely that MW tidal forces would play any major role at this location in its orbit; although, see \citet{Riley2024} and \citet{Shipp2024} for simulated streams that form on comparable orbits, including via preprocessing. 

Systems undergoing tidal disruption are expected to have moderately large ellipticities, but this is not a requirement \citep[e.g.][]{Munoz2008}. 
For example, Tucana~III \citep[$\epsilon=0.2\pm0.1$;][]{DW2015, Shipp2018, Burcin2018} and Crater~II \citep[$\epsilon=0.12\pm0.02$;][]{Torrealba2016a, Vivas2020, Ji2021} show clear signs of tidal disruption despite relatively small ellipticities. Therefore, Eri~IV's modest ellipticity ($\epsilon=\eriELLIP\pm\eriELLIPerr$) does not necessarily conflict with a potential tidal stripping scenario. Ursa Major~II \citep[UMa~II,][]{Zucker2006, Munoz2010, Munoz2018} occupies a similar space as Eri~IV on the size-luminosity plane and is thought to be tidally disrupting; although, UMa~II does has a high ellipticity ($\epsilon=0.56$). Eri~IV also resembles two MW UFDs in terms of its structural parameters: Pisces~II \citep{Belokurov2010, Sand2012, Richstein2022} and Coma Berenices \citep{Belokurov2007, Munoz2018}; neither of which show extended features to the same extent as Eri~IV. 
Hercules is a highly elongated UFD long thought to be tidally disrupting, yet conclusive evidence has remained elusive \citep[e.g.][]{Burcin2020}. 
\citet{Ou2024} recently showed that dynamical models of stream formation can replicate Hercules' morphology while indicating that any radial velocity gradient via tidal disruption would be undetectable given their sample size of 28 stars. 

Our photometric evidence suggests, with high confidence, that the northeast extension is a real feature. 
One explanation for the feature is that Eri~IV is being tidally disrupted, but the sample size of spectroscopic stars is too small to detect a velocity gradient, and/or our orbital assumptions are incorrect. 
However, tidal disruption is not the only explanation for the extended feature; this region could be interpreted as a stellar halo, captured field stars, or a satellite of Eri~IV. 
\citet{Chiti2021} found an extended component (member stars out to $9r_{h}$) of Tucana~II; suggestive of strong bursty feedback or an early merger. 
\citet{Tarumi2021} show that major mergers of UFDs at early times can produce extended stellar haloes similar to Tucana~II, and halo production is quite sensitive to the merger mass ratio \citep[e.g.][]{Querci2024}.
\citet{Penarrubia2024} show that stellar clumps, similar to the extension in Eri~IV, can be explained by starless dark matter sub-subhaloes capturing field stars. 
These clumps would contain stellar populations indistinguishable from the host galaxy; \citet{Heiger2023} found no evidence of a radial metallicity gradient in Eri~IV which would support this idea.
The feature may also be a sub-satellite of Eri~IV (i.e. satellite of a UFD), which is predicted by $\Lambda$CDM \citep[e.g.][]{Bullock2017}.
However, given there is no observational evidence of UFD-satellites or field star capture, we favor the tidal disruption or stellar halo scenarios. 
That said, we do not have sufficient evidence to pin down the exact nature of Eri~IV's extended feature.

\section{Summary and Conclusion} \label{sec:conclusion}

We present deep Magellan$+$Megacam photometric observations of two MW satellites: Cen~I and Eri~IV. These UFDs were recently discovered in the DELVE survey, and observed spectroscopically by \citet{Heiger2023}. Our data probe $\sim2-3$ magnitudes deeper than the discovery data,  allowing us to derive robust distance measurements, constrain the structural properties, luminosities, and find a handful of potential new member stars. We construct high-quality morphological maps, and perform extensive searches for extended structures to assess the dynamical state and nature of these systems. We find that Cen~I does not have a tidal feature associated with it, but Eri~IV hosts an extended feature with multiple interpretations. 

These deep data confirm that Cen~I aligns well with the discovery observations in terms of its distance, structural parameters, and luminosity. 
We find six potential new member stars based on \textit{Gaia} proper motions.
For Eri~IV, our data reveal it to be slightly closer ($D=\eriDKPC\pm\eriDKPCerr$ kpc), more round ($\epsilon=\eriELLIP\pm\eriELLIPerr$), and also dimmer, at $M_{V}=\eriMV\pm\eriMVerr$ mag. 
Aside from these differences, our measurements for Eri~IV are consistent with the discovery data in the following parameters: $\alpha_{\text{2000}}$, $\delta_{\text{2000}}$, $r_{h}$, and position angle. 
There are 14 potential new member stars based on \textit{Gaia} proper motions.
Both systems occupy a space on the size-luminosity plane consistent with other confirmed UFDs.

Our work helps to clear up some of the open questions surrounding the discovery photometry and the spectroscopic data: both Cen~I and Eri~IV exhibited hints of extended features in the discovery data, but spectroscopic analysis suggested these overdensities were not the result of tidal disruption. 
Our photometry shows no compelling evidence that Cen~I is disrupting.
Eri~IV, on the other hand, hosts an extended structure to the northeast of its central body (Figure \ref{fig:ext_struct}). 
This structure may be the result of tidal disruption in the recent past, although we can't rule out the possibility that this region arose from a merger or it is a satellite itself. 
It seems as though Eri~IV will remain an interesting UFD moving forward; the true nature of the extended structure may be understood through a combination of deeper and wider-field imaging data and detailed dynamical modeling.
This work highlights the importance of deep imaging in conjunction with spectroscopy to specifically explore the extended structures, ultimately aiding in the understanding of the true nature of ultra-faint satellites. 



\section{Acknowledgments}

We thank the anonymous reviewer whose comments and suggestions improved the content of this paper. 
QOC acknowledges support from the Dartmouth Fellowship. 
DJS and the Arizona team acknowledge support from NSF grant AST-2205863. QOC thanks Ryan Hickox and Elisabeth Newton for useful discussions. 

This paper includes data gathered with the 6.5-meter Magellan Telescope located at Las Campanas Observatory, Chile. 

The DECam Local Volume Exploration Survey (DELVE; NOAO Proposal ID 2019A-0305, PI: Drlica-Wagner) is partially supported by Fermilab LDRD project L2019-011 and the NASA Fermi Guest Investigator Program Cycle 9 No. 91201.
This project used data obtained with the Dark Energy Camera (DECam), which was constructed by the Dark Energy Survey (DES) collaboration. Funding for the DES Projects has been provided by the U.S. Department of Energy, the U.S. National Science Foundation, the Ministry of Science and Education of Spain, the Science and Technology Facilities Council of the United Kingdom, the Higher Education Funding Council for England, the National Center for Supercomputing Applications at the University of Illinois at Urbana–Champaign, the Kavli Institute of Cosmological Physics at the University of Chicago, the Center for Cosmology and Astro-Particle Physics at the Ohio State University, the Mitchell Institute for Fundamental Physics and Astronomy at Texas A\&M University, Financiadora de Estudos e Projetos, Fundação Carlos Chagas Filho de Amparo à Pesquisa do Estado do Rio de Janeiro, Conselho Nacional de Desenvolvimento Científico e Tecnológico and the Ministério da Ciência, Tecnologia e Inovação, the Deutsche Forschungsgemeinschaft and the Collaborating Institutions in the Dark Energy Survey.
The Collaborating Institutions are Argonne National Laboratory, the University of California at Santa Cruz, the University of Cambridge, Centro de Investigaciones Enérgeticas, Medioambientales y Tecnológicas–Madrid, the University of Chicago, University College London, the DES-Brazil Consortium, the University of Edinburgh, the Eidgenössische Technische Hochschule (ETH) Zürich, Fermi National Accelerator Laboratory, the University of Illinois at Urbana-Champaign, the Institut de Ciències de l'Espai (IEEC/CSIC), the Institut de Física d'Altes Energies, Lawrence Berkeley National Laboratory, the Ludwig-Maximilians Universität München and the associated Excellence Cluster Universe, the University of Michigan, the National Optical Astronomy Observatory, the University of Nottingham, the Ohio State University, the OzDES Membership Consortium, the University of Pennsylvania, the University of Portsmouth, SLAC National Accelerator Laboratory, Stanford University, the University of Sussex, and Texas A\&M University.
Based in part on observations at Cerro Tololo Inter-American Observatory, National Optical Astronomy Observatory, which is operated by the Association of Universities for Research in Astronomy (AURA) under a cooperative agreement with the National Science Foundation.
Database access and other data services are hosted by the Astro Data Lab at the Community Science and Data Center (CSDC) of the National Science Foundation's National Optical Infrared Astronomy Research Laboratory, operated by the Association of Universities for Research in Astronomy (AURA) under a cooperative agreement with the National Science Foundation.

This work has made use of data from the European Space Agency (ESA) mission
{\it Gaia} (\url{https://www.cosmos.esa.int/gaia}), processed by the {\it Gaia}
Data Processing and Analysis Consortium (DPAC,
\url{https://www.cosmos.esa.int/web/gaia/dpac/consortium}). Funding for the DPAC
has been provided by national institutions, in particular the institutions
participating in the {\it Gaia} Multilateral Agreement.

\hfil 

\textit{Facilities:} Las Campanas Observatory: Magellan Clay Telescope/Megacam, \textit{Gaia}

\textit{Software:} IDL astronomy users library \citep{Landsman1993}, SExtractor \citep{sextractor1996}, numpy \citep{numpy2020}, pandas \citep{pandas2021}, astropy \citep{astropy2022}, Topcat \citep{Taylor2005}.



\begin{thebibliography}{}


\bibitem[Astropy Collaboration et al.(2022)]{astropy2022} Astropy Collaboration, Price-Whelan, A.~M., Lim, P.~L., et al.\ 2022, \apj, 935, 167. doi:10.3847/1538-4357/ac7c74



\bibitem[Bechtol et al.(2015)]{Bechtol2015} Bechtol, K., Drlica-Wagner, A., Balbinot, E., et al.\ 2015, \apj, 807, 50. doi:10.1088/0004-637X/807/1/50

\bibitem[Belokurov et al.(2007)]{Belokurov2007} Belokurov, V., Zucker, D.~B., Evans, N.~W., et al.\ 2007, \apj, 654, 897. doi:10.1086/509718

\bibitem[Belokurov et al.(2010)]{Belokurov2010} Belokurov, V., Walker, M.~G., Evans, N.~W., et al.\ 2010, \apjl, 712, L103. doi:10.1088/2041-8205/712/1/L103


\bibitem[Bernard et al.(2014)]{Bernard2014} Bernard, E.~J., Ferguson, A.~M.~N., Schlafly, E.~F., et al.\ 2014, \mnras, 442, 2999. doi:10.1093/mnras/stu1081

\bibitem[Bertin \& Arnouts(1996)]{sextractor1996} Bertin, E. \& Arnouts, S.\ 1996, \aaps, 117, 393. doi:10.1051/aas:1996164

\bibitem[Bertin et al.(2002)]{Bertin2002} Bertin, E., Mellier, Y., Radovich, M., et al.\ 2002, Astronomical Data Analysis Software and Systems XI, 281, 228

\bibitem[Binney \& Tremaine(2008)]{Binney2008} Binney, J. \& Tremaine, S.\ 2008, Galactic Dynamics: Second Edition, by James Binney and Scott Tremaine. ISBN 978-0-691-13026-2 (HB). Published by Princeton University Press, Princeton, NJ USA, 2008.




\bibitem[Bressan et al.(2012)]{Bressen2012} Bressan, A., Marigo, P., Girardi, L., et al.\ 2012, \mnras, 427, 127. doi:10.1111/j.1365-2966.2012.21948.x

\bibitem[Bullock \& Boylan-Kolchin(2017)]{Bullock2017} Bullock, J.~S. \& Boylan-Kolchin, M.\ 2017, \araa, 55, 343. doi:10.1146/annurev-astro-091916-055313

\bibitem[Cantu et al.(2021)]{Cantu2021} Cantu, S.~A., Pace, A.~B., Marshall, J., et al.\ 2021, \apj, 916, 81. doi:10.3847/1538-4357/ac0443

\bibitem[Carlin et al.(2017)]{Carlin2017} Carlin, J.~L., Sand, D.~J., Mu{\~n}oz, R.~R., et al.\ 2017, \aj, 154, 267. doi:10.3847/1538-3881/aa94d0


\bibitem[Cerny et al.(2021)]{Cerny2021} Cerny, W., Pace, A.~B., Drlica-Wagner, A., et al.\ 2021, \apjl, 920, L44. doi:10.3847/2041-8213/ac2d9a

\bibitem[Cerny et al.(2023a)]{Cerny2023a} Cerny, W., Simon, J.~D., Li, T.~S., et al.\ 2023, \apj, 942, 111. doi:10.3847/1538-4357/aca1c3

\bibitem[Cerny et al.(2023b)]{Cerny2023b} Cerny, W., Mart{\'\i}nez-V{\'a}zquez, C.~E., Drlica-Wagner, A., et al.\ 2023, \apj, 953, 1. doi:10.3847/1538-4357/acdd78

\bibitem[Cerny et al.(2023c)]{Cerny2023c} Cerny, W., Drlica-Wagner, A., Li, T.~S., et al.\ 2023, \apjl, 953, L21. doi:10.3847/2041-8213/aced84

\bibitem[Cerny et al.(2024)]{Cerny2024} Cerny, W., Chiti, A., Geha, M., et al.\ 2024, arXiv:2410.00981. doi:10.48550/arXiv.2410.00981



\bibitem[Chiti et al.(2021)]{Chiti2021} Chiti, A., Frebel, A., Simon, J.~D., et al.\ 2021, Nature Astronomy, 5, 392. doi:10.1038/s41550-020-01285-w


\bibitem[Choi et al.(2018)]{Choi2018} Choi, Y., Nidever, D.~L., Olsen, K., et al.\ 2018, \apj, 869, 125. doi:10.3847/1538-4357/aaed1f


\bibitem[Crnojevi{\'c} et al.(2016)]{Crnojevic2016} Crnojevi{\'c}, D., Sand, D.~J., Zaritsky, D., et al.\ 2016, \apjl, 824, L14. doi:10.3847/2041-8205/824/1/L14





\bibitem[Drlica-Wagner et al.(2015)]{DW2015} Drlica-Wagner, A., Bechtol, K., Rykoff, E.~S., et al.\ 2015, \apj, 813, 109. doi:10.1088/0004-637X/813/2/109

\bibitem[Drlica-Wagner et al.(2016)]{DW2016} Drlica-Wagner, A., Bechtol, K., Allam, S., et al.\ 2016, \apjl, 833, L5. doi:10.3847/2041-8205/833/1/L5

\bibitem[Drlica-Wagner et al.(2022)]{delvedr2} Drlica-Wagner, A., Ferguson, P.~S., Adam{\'o}w, M., et al.\ 2022, \apjs, 261, 38. doi:10.3847/1538-4365/ac78eb

\bibitem[Dotter et al.(2008)]{Dotter2008} Dotter, A., Chaboyer, B., Jevremovi{\'c}, D., et al.\ 2008, \apjs, 178, 89. doi:10.1086/589654

\bibitem[Eyer et al.(2023)]{Eyer2023} Eyer, L., Audard, M., Holl, B., et al.\ 2023, \aap, 674, A13. doi:10.1051/0004-6361/202244242



\bibitem[Gaia Collaboration et al.(2016)]{Gaia2016} Gaia Collaboration, Prusti, T., de Bruijne, J.~H.~J., et al.\ 2016, \aap, 595, A1. doi:10.1051/0004-6361/201629272


\bibitem[Gaia Collaboration et al.(2021)]{Gaia2021} Gaia Collaboration, Brown, A.~G.~A., Vallenari, A., et al.\ 2021, \aap, 649, A1. doi:10.1051/0004-6361/202039657



\bibitem[Gonz{\'a}lez-Morales et al.(2017)]{Gonzalez2017} Gonz{\'a}lez-Morales, A.~X., Marsh, D.~J.~E., Pe{\~n}arrubia, J., et al.\ 2017, \mnras, 472, 1346. doi:10.1093/mnras/stx1941


\bibitem[Harris(2010)]{Harris2010} Harris, W.~E.\ 2010, arXiv:1012.3224. doi:10.48550/arXiv.1012.3224

\bibitem[Harris et al.(2020)]{numpy2020}Harris, C.R., Millman, K.J., van der Walt, S.J. et al.\ 2020, \nat, 585, 357. doi:10.1038/s41586-020-2649-2

\bibitem[Heiger et al.(2024)]{Heiger2023} Heiger, M.~E., Li, T.~S., Pace, A.~B., et al.\ 2024, \apj, 961, 234. doi:10.3847/1538-4357/ad0cf7

\bibitem[Homma et al.(2018)]{Homma2018} Homma, D., Chiba, M., Okamoto, S., et al.\ 2018, \pasj, 70, S18. doi:10.1093/pasj/psx050

\bibitem[Homma et al.(2019)]{Homma2019} Homma, D., Chiba, M., Komiyama, Y., et al.\ 2019, \pasj, 71, 94. doi:10.1093/pasj/psz076

\bibitem[Homma et al.(2023)]{Homma2023} Homma, D., Chiba, M., Komiyama, Y., et al.\ 2023, arXiv:2311.05439. doi:10.48550/arXiv.2311.05439



\bibitem[Ji et al.(2021)]{Ji2021} Ji, A.~P., Koposov, S.~E., Li, T.~S., et al.\ 2021, \apj, 921, 32. doi:10.3847/1538-4357/ac1869

\bibitem[Kim \& Jerjen(2015)]{Kim2015} Kim, D. \& Jerjen, H.\ 2015, \apjl, 808, L39. doi:10.1088/2041-8205/808/2/L39



\bibitem[Koposov et al.(2015)]{Koposov2015} Koposov, S.~E., Belokurov, V., Torrealba, G., et al.\ 2015, \apj, 805, 130. doi:10.1088/0004-637X/805/2/130

\bibitem[Koposov et al.(2018)]{Koposov2018} Koposov, S.~E., Walker, M.~G., Belokurov, V., et al.\ 2018, \mnras, 479, 5343. doi:10.1093/mnras/sty1772



\bibitem[Landsman(1993)]{Landsman1993} Landsman, W.~B.\ 1993, Astronomical Data Analysis Software and Systems II, 52, 246





\bibitem[Martin et al.(2008)]{Martin2008} Martin, N.~F., de Jong, J.~T.~A., \& Rix, H.-W.\ 2008, \apj, 684, 1075. doi:10.1086/590336


\bibitem[Martin et al.(2015)]{Martin2015} Martin, N.~F., Nidever, D.~L., Besla, G., et al.\ 2015, \apjl, 804, L5. doi:10.1088/2041-8205/804/1/L5

\bibitem[Martin et al.(2016)]{Martin2016} Martin, N.~F., Ibata, R.~A., Lewis, G.~F., et al.\ 2016, \apj, 833, 167. doi:10.3847/1538-4357/833/2/167

\bibitem[Mart{\'\i}nez-V{\'a}zquez et al.(2019)]{Martinez-Vazquez2019} Mart{\'\i}nez-V{\'a}zquez, C.~E., Vivas, A.~K., Gurevich, M., et al.\ 2019, \mnras, 490, 2183. doi:10.1093/mnras/stz2609


\bibitem[Mart{\'\i}nez-V{\'a}zquez et al.(2021)]{Martinez-Vazquez2021} Mart{\'\i}nez-V{\'a}zquez, C.~E., Cerny, W., Vivas, A.~K., et al.\ 2021, \aj, 162, 253. doi:10.3847/1538-3881/ac2368

\bibitem[Mau et al.(2019)]{Mau2019} Mau, S., Drlica-Wagner, A., Bechtol, K., et al.\ 2019, \apj, 875, 154. doi:10.3847/1538-4357/ab0bb8

\bibitem[Mau et al.(2020)]{Mau2020} Mau, S., Cerny, W., Pace, A.~B., et al.\ 2020, \apj, 890, 136. doi:10.3847/1538-4357/ab6c67

\bibitem[McConnachie(2012)]{McConnachie2012} McConnachie, A.~W.\ 2012, \aj, 144, 4. doi:10.1088/0004-6256/144/1/4



\bibitem[McLeod et al.(2015)]{McLeod2015} McLeod, B., Geary, J., Conroy, M., et al.\ 2015, \pasp, 127, 366. doi:10.1086/680687

\bibitem[McNanna et al.(2024)]{McNanna2024} McNanna, M., Bechtol, K., Mau, S., et al.\ 2024, \apj, 961, 126. doi:10.3847/1538-4357/ad07d0


\bibitem[Moskowitz \& Walker(2020)]{Moskowitz2020} Moskowitz, A.~G. \& Walker, M.~G.\ 2020, \apj, 892, 27. doi:10.3847/1538-4357/ab7459

\bibitem[Mu{\~n}oz et al.(2008)]{Munoz2008} Mu{\~n}oz, R.~R., Majewski, S.~R., \& Johnston, K.~V.\ 2008, \apj, 679, 346. doi:10.1086/587125

\bibitem[Mu{\~n}oz et al.(2010)]{Munoz2010} Mu{\~n}oz, R.~R., Geha, M., \& Willman, B.\ 2010, \aj, 140, 138. doi:10.1088/0004-6256/140/1/138

\bibitem[Mu{\~n}oz et al.(2018)]{Munoz2018} Mu{\~n}oz, R.~R., C{\^o}t{\'e}, P., Santana, F.~A., et al.\ 2018, \apj, 860, 66. doi:10.3847/1538-4357/aac16b

\bibitem[Mutlu-Pakdil et al.(2018)]{Burcin2018} Mutlu-Pakdil, B., Sand, D.~J., Carlin, J.~L., et al.\ 2018, \apj, 863, 25. doi:10.3847/1538-4357/aacd0e

\bibitem[Mutlu-Pakdil et al.(2020)]{Burcin2020} Mutlu-Pakdil, B., Sand, D.~J., Crnojevi{\'c}, D., et al.\ 2020, \apj, 902, 106. doi:10.3847/1538-4357/abb40b





\bibitem[Ou et al.(2024)]{Ou2024} Ou, X., Chiti, A., Shipp, N., et al.\ 2024, \apj, 966, 33. doi:10.3847/1538-4357/ad2f27

\bibitem[Pace \& Li(2019)]{Pace2019} Pace, A.~B. \& Li, T.~S.\ 2019, \apj, 875, 77. doi:10.3847/1538-4357/ab0aee

\bibitem[Pace et al.(2022)]{Pace2022} Pace, A.~B., Erkal, D., \& Li, T.~S.\ 2022, \apj, 940, 136. doi:10.3847/1538-4357/ac997b

\bibitem[Pace(2024)]{Pace2024} Pace, A.~B.\ 2024, arXiv:2411.07424




\bibitem[Pe{\~n}arrubia et al.(2024)]{Penarrubia2024} Pe{\~n}arrubia, J., Errani, R., Walker, M.~G., et al.\ 2024, \mnras, 533, 3263. doi:10.1093/mnras/stae1961

\bibitem[Pietrinferni et al.(2004)]{Pietrinferni2004} Pietrinferni, A., Cassisi, S., Salaris, M., et al.\ 2004, \apj, 612, 168. doi:10.1086/422498

\bibitem[Querci et al.(2024)]{Querci2024} Querci, L., Pallottini, A., Branca, L., et al.\ 2024, arXiv:2410.05396. doi:10.48550/arXiv.2410.05396


\bibitem[Reback et al.(2021)]{pandas2021} Reback, J., Jbrockmendel, McKinney, W., et al.\ 2021, Zenodo


\bibitem[Rhode et al.(2023)]{Rhode2023} Rhode, K.~L., Smith, N.~J., Crnojevic, D., et al.\ 2023, \aj, 166, 180. doi:10.3847/1538-3881/acf859

\bibitem[Richstein et al.(2022)]{Richstein2022} Richstein, H., Patel, E., Kallivayalil, N., et al.\ 2022, \apj, 933, 217. doi:10.3847/1538-4357/ac7226

\bibitem[Riley et al.(2024)]{Riley2024} Riley, A.~H., Shipp, N., Simpson, C.~M., et al.\ 2024, arXiv:2410.09144. doi:10.48550/arXiv.2410.09144

\bibitem[Rockosi et al.(2002)]{Rockosi2002} Rockosi, C.~M., Odenkirchen, M., Grebel, E.~K., et al.\ 2002, \aj, 124, 349. doi:10.1086/340957

\bibitem[Safarzadeh \& Spergel(2020)]{Safarzadeh2020} Safarzadeh, M. \& Spergel, D.~N.\ 2020, \apj, 893, 21. doi:10.3847/1538-4357/ab7db2

\bibitem[Salpeter(1955)]{Salpeter1955} Salpeter, E.~E.\ 1955, \apj, 121, 161. doi:10.1086/145971

\bibitem[Sand et al.(2009)]{Sand2009} Sand, D.~J., Olszewski, E.~W., Willman, B., et al.\ 2009, \apj, 704, 898. doi:10.1088/0004-637X/704/2/898

\bibitem[Sand et al.(2010)]{Sand2010} Sand, D.~J., Seth, A., Olszewski, E.~W., et al.\ 2010, \apj, 718, 530. doi:10.1088/0004-637X/718/1/530

\bibitem[Sand et al.(2012)]{Sand2012} Sand, D.~J., Strader, J., Willman, B., et al.\ 2012, \apj, 756, 79. doi:10.1088/0004-637X/756/1/79




\bibitem[Schlegel et al.(1998)]{Schlegel1998} Schlegel, D.~J., Finkbeiner, D.~P., \& Davis, M.\ 1998, \apj, 500, 525. doi:10.1086/305772

\bibitem[Shipp et al.(2018)]{Shipp2018} Shipp, N., Drlica-Wagner, A., Balbinot, E., et al.\ 2018, \apj, 862, 114. doi:10.3847/1538-4357/aacdab

\bibitem[Shipp et al.(2023)]{Shipp2023} Shipp, N., Panithanpaisal, N., Necib, L., et al.\ 2023, \apj, 949, 44. doi:10.3847/1538-4357/acc582

\bibitem[Shipp et al.(2024)]{Shipp2024} Shipp, N., Riley, A.~H., Simpson, C.~M., et al.\ 2024, arXiv:2410.09143. doi:10.48550/arXiv.2410.09143

\bibitem[Simon(2019)]{Simon2019} Simon, J.~D.\ 2019, \araa, 57, 375. doi:10.1146/annurev-astro-091918-104453

\bibitem[Simon et al.(2020)]{Simon2020} Simon, J.~D., Li, T.~S., Erkal, D., et al.\ 2020, \apj, 892, 137. doi:10.3847/1538-4357/ab7ccb

\bibitem[Smith et al.(2023)]{Smith2023} Smith, S.~E.~T., Jensen, J., Roediger, J., et al.\ 2023, \aj, 166, 76. doi:10.3847/1538-3881/acdd77

\bibitem[Stetson(1994)]{Stetson1994} Stetson, P.~B.\ 1994, \pasp, 106, 250. doi:10.1086/133378

\bibitem[Strigari(2018)]{Strigari2018} Strigari, L.~E.\ 2018, Reports on Progress in Physics, 81, 056901. doi:10.1088/1361-6633/aaae16

\bibitem[Tan et al.(2024)]{Tan2024} Tan, C.~Y., Cerny, W., Drlica-Wagner, A., et al.\ 2024, arXiv:2408.00865. doi:10.48550/arXiv.2408.00865


\bibitem[Tarumi et al.(2021)]{Tarumi2021} Tarumi, Y., Yoshida, N., \& Frebel, A.\ 2021, \apjl, 914, L10. doi:10.3847/2041-8213/ac024e

\bibitem[Taylor(2005)]{Taylor2005} Taylor, M.~B.\ 2005, Astronomical Data Analysis Software and Systems XIV, 347, 29

\bibitem[Tollerud et al.(2012)]{Tollerud2012} Tollerud, E.~J., Beaton, R.~L., Geha, M.~C., et al.\ 2012, \apj, 752, 45. doi:10.1088/0004-637X/752/1/45

\bibitem[Torrealba et al.(2016a)]{Torrealba2016a} Torrealba, G., Koposov, S.~E., Belokurov, V., et al.\ 2016, \mnras, 459, 2370. doi:10.1093/mnras/stw733

\bibitem[Torrealba et al.(2016b)]{Torrealba2016b} Torrealba, G., Koposov, S.~E., Belokurov, V., et al.\ 2016, \mnras, 463, 712. doi:10.1093/mnras/stw2051

\bibitem[Torrealba et al.(2018)]{Torrealba2018} Torrealba, G., Belokurov, V., Koposov, S.~E., et al.\ 2018, \mnras, 475, 5085. doi:10.1093/mnras/sty170

\bibitem[Vivas et al.(2020)]{Vivas2020} Vivas, A.~K., Walker, A.~R., Mart{\'\i}nez-V{\'a}zquez, C.~E., et al.\ 2020, \mnras, 492, 1061. doi:10.1093/mnras/stz3393

\bibitem[Walsh et al.(2008)]{Walsh2008} Walsh, S.~M., Willman, B., Sand, D., et al.\ 2008, \apj, 688, 245. doi:10.1086/592076

\bibitem[Wang et al.(2019)]{Wang2019} Wang, M.~Y., de Boer, T., Pieres, A., et al.\ 2019, \apj, 881, 118. doi:10.3847/1538-4357/ab31a9

\bibitem[Weinberg et al.(2015)]{Weinberg2015} Weinberg, D.~H., Bullock, J.~S., Governato, F., et al.\ 2015, Proceedings of the National Academy of Science, 112, 12249. doi:10.1073/pnas.1308716112


\bibitem[Wetzel et al.(2016)]{Wetzel2016} Wetzel, A.~R., Hopkins, P.~F., Kim, J.-. hoon ., et al.\ 2016, \apjl, 827, L23. doi:10.3847/2041-8205/827/2/L23

\bibitem[Zucker et al.(2006)]{Zucker2006} Zucker, D.~B., Belokurov, V., Evans, N.~W., et al.\ 2006, \apjl, 650, L41. doi:10.1086/508628

\end{thebibliography}
\end{document}